\documentclass{aa}  
\usepackage{natbib}
\bibpunct{(}{)}{;}{a}{}{,} 
\usepackage{graphicx}
\usepackage{txfonts}
\begin{document} 

   \title{Supernova explosions of runaway stars and young neutron stars above the Galactic plane}


   \author{V.~Daki\'{c}
          \inst{1} \and S. B.~Popov\inst{2}
          \and R.~Turolla\inst{1,}\inst{3}}

   \institute{Dipartimento di Fisica e Astronomia “G.~Galilei”, Università di Padova, Via Marzolo 8, 35121 Padova, Italy
         \and
           Sternberg Astronomical Institute, Lomonosov Moscow State University, Universitetsky prospekt 13, 119234 Moscow, Russia
           \and
           Mullard Space Science Laboratory, University College London, Holmbury St Mary, Dorking, Surrey RH5 6NT, UK
             }

   \date{Received \today \ Accepted <date>}

 
  \abstract
   {Several supernova remnants and young neutron stars were recently discovered relatively high above the Galactic plane. One possibility is that they originate from runaway OB stars born in the Galactic disk. Understanding their origin will provide key insights into the properties of the Galactic halo. }
   {This paper aims to determine the spatial distribution of supernova explosions from runaway OB stars and to assess whether this model can explain certain observed neutron stars.}
   {We map the distribution of supernova events produced by runaway OB stars by incorporating their birth rate, initial spatial distribution, lifetime, ejection mechanisms, and velocity distributions. By tracking their motion in the Galactic potential, we determine their final distribution right before the explosion.}
   {
   We show that the neutron star Calvera, which is found at $z\approx2.2~\mathrm{kpc}$, could have originated from a runaway OB star. In addition, we compare the probabilities of finding a supernova remnant originating from Type Ia and core-collapse supernovae high above the Galactic plane, showing that supernova remnants related to core-collapse supernovae outnumber those related to Type Ia supernovae.
}
   {}

   \keywords{stars: neutron --
   stars: early-type --
   stars: kinematics and dynamics --
   stars: massive --
   supernovae: general}

   \maketitle
%
\section{Introduction}
Star formation in the Milky Way (MW) is believed to be confined to the Galactic disk, mostly in the so-called star-forming regions. Thus, the youngest stellar population is mainly observed in structures such as OB associations and open clusters. In the photometric survey by \cite{1947ApJ...105...85H}, faint blue stars at high Galactic latitudes have been identified. They are low-mass, evolved objects that belong to Population II stars. Further studies have shown that main-sequence (MS) stars are also present in the Galactic halo, although they appear very faint due to the large distances \citep{1974ApJS...28..157G}. Later, the inherent difficulty in disentangling the latter population from the former was appreciated \citep{1987fbs..conf.....P}, thus analyzing the stellar parameters and element abundances of faint blue stars at high Galactic latitudes is crucial to determine whether they are MS stars. 

Young, blue early-type MS stars observed outside star-forming regions and having different kinematical properties compared to typical early-type MS stars are known as runaway stars. They are expected to be formed in the Galactic disk and later ejected to the Galactic halo. Using recent \textit{Gaia} DR3 astrometric data, \cite{2023OBper} estimated that $\approx 25\%$ of O-type stars and $\approx 5\%$ of Be-type stars are runaway stars. 
Two ejection mechanisms were proposed: ejection from a disrupted binary system and dynamical ejection due to stellar interactions in a dense environment \cite[see][]{2011Silva}. 
The binary ejection mechanism (BEM) was first proposed by \cite{1957moas.book.....Z} and later developed by \cite{1961BAN....15..265B}. In this scenario, the more massive component of a close binary system undergoes a supernova (SN) explosion. Due to mass loss or/and the kick imparted to the newborn compact object, the two stars become gravitationally unbound. The secondary component has a large orbital velocity which is mostly conserved after the binary is disrupted and becomes a runaway star. 
In its original form, this model predicts that runaway stars can appear only as single objects. From simulations, it is expected that the maximum ejection velocity is around $300\ \mathrm{km\,s^{-1}}$, or even up to $400\ \mathrm{km\,s^{-1}}$ \citep{2000ApJ...544..437P}. 

The dynamical ejection mechanism (DEM) was proposed by \cite{1967BOTT....4...86P} as an alternative to the BEM. In this scenario, runaway stars are ejected as a result of dynamical interactions due to gravitational instabilities during the early phases of an open cluster evolution. The dynamic interaction between the stars in the young cluster could result in the ejection of both single and multiple stars. Runaway stars ejected this way are expected to have a velocity of up to $400~\mathrm{km\,s^{-1}}$ \citep{2009MNRAS.395L..85G, 2009MNRAS.396..570G}. 

By tracing the orbits of runaway stars back to their parent clusters, it has been speculated that both mechanisms may be at work, contributing roughly equally to the population of runaway stars \citep{2001A&A...365...49H}. This conclusion, however, was based on a very limited number of 
representative cases. In particular, the association of $\zeta~\mathrm{Oph}$ with a nearby pulsar as evidence of a BEM was later refuted by \cite{2020MNRAS.498..899N}, who showed that the two objects do not share a common origin. This finding, supported by subsequent studies, undermines the assumption of equal contributions from BEM and DEM and highlights the need to reevaluate the relative importance of these mechanisms using more robust statistical samples. Still, this conclusion was confirmed in recent studies on the origin of B runaway stars, analyzing their chemical composition and kinematics \citep{2023Liu}. The bimodal distribution of ejection velocities observed in B-type stars suggests two distinct origins, aligning with the findings of \cite{2011Silva}. However, for runaway stars in the Small Magellanic Cloud (SMC), it was shown that the DEM dominates over the BEM by a factor $\sim2$--$3$ \citep{2018ApJ...867L...8O, 2020ApJ...903...43D}. 

The predicted ejection velocity distributions are similar for both mechanisms. \cite{2011Silva} proposed that there are two different populations of runaway stars: a high-velocity population with a maximum ejection velocity $\sim (400-500)~\mathrm{km\,s^{-1}}$, and a low-velocity population with a maximum ejection velocity $\sim 300\,\mathrm{km\,s^{-1}}$. These authors argue that the observed limit of $\sim 500\, \mathrm{km\, s^{-1}}$ and the bimodality of the observed distribution are natural consequences of the BEM. In addition, \cite{2011Silva} found three stars that were inconsistent with the ejection from the Galactic disk, discussing different possible scenarios, including a possible formation in the Galactic halo. 

Actually, the BEM and DEM may act jointly, resulting in the so-called two-step ejection mechanism, which was proposed as a natural explanation for the presence of apparently isolated massive stars \citep{2010MNRAS.404.1564P}. In this scenario, a binary system is first ejected via DEM and, subsequently, one of the binary components is ejected via BEM. This mechanism can produce stars that are indistinguishable from isolated field stars, thereby complicating the efforts to trace massive stars back to their birthplace and challenging the interpretation of isolated massive star formation as being truly in situ. A further study by \cite{2012MNRAS.424.3037G} showed that most of the known O stars thought to have formed in isolation are rather very likely runaway stars. Also, it was demonstrated that the field population includes O stars whose low space velocities and/or young ages appear inconsistent with the large distances separating them from their potential birth clusters or with the ages of those clusters. These stars, likely ejected by BEM, cannot be reliably traced back to their origins and may therefore be mistakenly classified as having formed in situ. In general, \cite{2012MNRAS.424.3037G} do not find compelling evidence supporting the in situ formation scenario for massive stars.

In addition, a recent study by \cite{2024A&A...690A.207P} suggests a possible third mechanism for the formation of runaway stars --- the subcluster ejection scenario (SCES). In this model, a subset of stars from an infalling subcluster is ejected out of the cluster via a tidal interaction with the contracting gravitational potential of the assembling cluster. Stars ejected via SCES share similar velocities, directions, and ages, and their ejection directions tend to be anisotropic, often aligned with the subcluster mergers geometry. It could explain grouped runaway stars that appear to be coeval and moving in similar directions (\citealt{2024Natur.634..809S}).

During their evolution, runaway OB stars can undergo a core-collapse supernova (CCSN) explosion, 
likely resulting in the formation of a compact object --- a neutron star (NS) or a black hole (BH). It is important to study the possible spatial distribution of these explosions, as several SNRs were recently observed high above the Galactic plane \cite[see e.g.][]{2015ApJ...812...37F, 2021Churazov}, all listed in the new catalog by \cite{2025Green}. Thanks to the SGR/eROSITA all-sky survey, a new X-ray-selected SNR candidate SRGe J0023+3625/G116.6-26.1 high above the Galactic plane was discovered \citep{2021Churazov}. It is hypothesized that this SNR originated from a Type Ia supernova (SN~Ia). Objects with parameters similar to those of G116.6-26.1 are rare, but not to the extent of making the association with G116.6-26.1 implausible \citep{2021Churazov}. However, other interpretations are possible.

In addition, young isolated NSs are also observed high above the Galactic plane. The most well-known example is the source 1RXS J141256.0+792204, dubbed Calvera. It has been first identified in the ROSAT All-Sky Survey Bright Source Catalog at Galactic latitude $b=+37^\circ$ \citep{2008ApJ...672.1137R}. Calvera is located at a distance of $\approx 3.3 ~\mathrm{kpc}$ and given its Galactic coordinates, this corresponds to a height of $\approx$2 kpc above the Galactic plane \citep{2021ApJ...922..253M}. Assuming that its characteristic age $\tau_c = 285 ~\mathrm{kyr}$ is the true age \citep{2013ApJ...778..120H}, it is possible to estimate the component of its velocity perpendicular to the Galactic plane. 
If the NS was born in the Galactic disk, it is $v_\mathrm{t} \approx 6800 ~\mathrm{km\,s^{-1}}$. This is unreasonably high compared to known kick velocities \citep{2024ApJ...976..228R}. In addition, Calvera's age is still uncertain. The proper motion measurements allowed \cite{2024ApJ...976..228R} to estimate the age as $<10~\mathrm{kyr}$. The SNR G118.4+37.0 is associated with Calvera, suggesting that Calvera originated from the same SN. The estimated age of the SNR is also $<10~\mathrm{kyr}$ \citep{2022A&A...667A..71A, 2023MNRAS.518.4132A}. 
There is a possibility that Calvera is a descendant of a runaway star \citep{2008A&A...482..617P}. Thus, modeling the population of runaway stars until they end up in a SN event might shed light on the origin of Calvera and similar sources.  

In this paper, we present a statistical analysis of runaway OB stars and calculate the rate of their explosions as SN. We modeled the distribution of SNRs produced by explosions of runaway OB stars relative to the Galactic plane, comparing it with that obtained by \cite{2024Bisht}. These authors analyze the possibility that the observed large column density absorption lines from a highly ionized gas can be explained by the reverse-shocked gas in non-radiative SNRs above the Galactic plane.
Such SNRs should originate from runaway stars.
We also analyze the probability that Calvera could originate from a runaway OB star. Finally, we compare the probabilities of finding an SNR formed after a CCSN explosion high above the Galactic plane and after an SN~Ia following the method presented by \cite{2021Churazov}. 

The paper is organized as follows. In Sect.~\ref{Model}, we describe our assumptions and methods for modeling the population of runaway OB stars and tracking their motion in the Galactic potential. In Sect.~\ref{Results}, we present the spatial distribution of SNRs originating from the collapse of runaway OB stars, analyze the case of Calvera, and compare the distributions of CCSN and SN~Ia high above the Galactic plane. Discussion and conclusions follow in Sect.~\ref{Discussion} and Sect.~\ref{Conclusion}. 

\section{Model} 
\label{Model}

In this section, we present the model used to calculate the spatial distribution of SN explosions of runaway OB stars. First, we present the parameters of OB stars and then describe how we calculate their trajectories in the Galactic potential.
\subsection{Parameters of OB stars}
\label{Initial parameters}

There are four main ingredients of our model:
the mass-dependent birth rate of OB stars, their initial spatial distribution, the velocity distribution, and, finally, the age at which a star is ejected due to binary disruption or to dynamical interaction.

\subsubsection{Birth rate}
\label{birth}
The distribution of stellar masses in a given stellar population follows a statistical distribution known as the initial mass function (IMF), which is defined in terms of the number of stars per logarithmic mass interval, $\xi(\log M)={dN}/{d\log M}$. We adopted the Chabrier IMF for single stars in the Galactic disk \citep{Chabrier03}. The Chabrier IMF consists of a log-normal distribution at low stellar masses ($M\leq1M_\sun$):
\begin{equation}
\xi(\log M/M_\sun)_{M\leq1M_\sun} =
A\exp\left[ -\dfrac{(\log (M/M_\sun) - \log M_c)^2}{2\sigma^2} \right]
\label{eq:m<1}
\end{equation}
and a power-law tail at higher masses ($M>1M_\sun$):
\begin{equation}
\xi(\log M/M_\sun)_{M>1M_\sun} = BM^{-x}
\label{eq:m>1}
\end{equation}
where $A=0.158^{+0.051}_{-0.046}$, $M_c \approx 0.079^{-0.016}_{+0.021}$, $\sigma \approx 0.69^{-0.01}_{+0.05}$, $B=4.43\times 10^{-2}$, $x\approx 1.3\pm0.3$. The parameters $M_c, \sigma$, and $\alpha$ are empirically determined based on observational data, mainly from stellar counts and luminosity functions in the solar neighborhood, while the constants $A$ and $B$ ensure continuity and proper normalization \cite[see][for details]{Chabrier03}. 

As we are interested in OB stars that undergo CCSN explosions, the range $M/M_\sun < 1$ can be ignored. Therefore, we proceed with the inverse transfer sampling of an arbitrary number of stars from Eq.~(\ref{eq:m>1}). Then, we select only stars with $M>8 M_\sun$ as the less massive stars will not undergo a CCSN event \citep{2015PASA...32...16S}. 

 After sampling $N_\mathrm{stars}$ stars in the mass range $8\, M_\sun<M<55 \, M_\sun$, we check which of them qualify as runaway stars. As discussed, for example, in \cite{2023OBper} the percentage of runaway stars depends on the spectral type, and the spectral type can be related to the mass \cite[see Table \ref{tab:ob};][]{2010A&A...524A..98W, 2014A&A...566A...7N}. Here, we accept the percentage of runaway B stars to be the same as that one of Be stars, as it is in agreement with the general assumption that 5-10\% of B stars are runaway stars and close to the value of 4\% proposed by \cite{1991AJ....102..333S}.
 
 For each bin, the central mass value is determined and the corresponding mass percentage is assigned to it. Then, linear interpolation is applied to obtain a continuous trend. Based on the probability obtained from the interpolation, for each star it is checked whether it is a runaway, obtaining the number of runaway stars ($N_\mathrm{runaway}$). Due to interpolation, the upper limit of the mass in our subset of runaway stars is $55 M_\sun$. However, the results are not sensitive to this as the number of massive stars becomes lower and lower at higher masses.

\begin{table}
\caption{The percentage of runaway stars depending on the spectral types and masses.}
\label{table:ob}
\centering
\begin{tabular}{c c c c}
\hline\hline
Spectral Type & Mass ($M_\sun$) & Runaway Star  & Refs. \\
 & & percentage (\%) & \\
\hline
    O2 - O7 & 25 - 86 & 25.1 & 1, 2\\ 
    O8 - O9 & 18 - 25 & 23.7 & 1, 2\\
    B0e - B3e & 8 - 18 & 6.2 & 1, 3 \\
    B4e - B9e & 2 - 8 & 4.8 & 1, 3 \\
    \hline 
    O8 - B1e & 15 - 25 & 14.8 & 1, 2, 3 \\
\hline 
    \end{tabular}
    \tablebib{(1) \citet{2023OBper}; (2) \citet{2010A&A...524A..98W}; (3) \citet{2014A&A...566A...7N}.}
    \label{tab:ob}
\end{table}

\subsubsection{Initial spatial distribution}
\label{spatial}

We assume that the initial spatial distribution of runaway OB stars has cylindrical symmetry, i.e. it is independent of the presence of arms and other spatial features of the Galaxy but depends only on the radial distance from the Galactic center, $R$, and the height above the Galactic plane, $z$. 
We also assume that the distribution of OB stars is similar to the distribution of young radio pulsars, which is well-traced in a relatively large volume of the Galaxy. 
Then the initial spatial density of OB stars is \citep{2006MNRAS.372..777L, 2016PhRvD..93a3009A}:
\begin{equation}
    \rho(R, z) \propto \left( \frac{R}{R_\sun} \right)^{\alpha} \exp \left( -\beta \frac{R - R_\sun}{R_\sun} \right) \exp \left( -\frac{|z|}{h} \right);
    \label{eq:spatial}
\end{equation}
here $R_\sun= 8.2~\mathrm{kpc}$ is the distance of the Sun from the Galactic center and $h= 0.181$ kpc is the vertical height scale. The coefficients $\alpha$ and $\beta$ are taken to be $\alpha = 1.93,\ \beta = 5.06$ \citep{2006MNRAS.372..777L, 2016PhRvD..93a3009A}. We proceed with the rejection sampling of $R$ and the inverse transform sampling of $z$ for each runaway star from Eq.~(\ref{eq:spatial}). The azimuthal angle $\varphi$ is randomly sampled, reflecting the assumed axisymmetry of the distribution.

\subsubsection{Velocity distribution}

We assume that before the ejection (acceleration), a star has the typical velocity of a massive MS star in the Galactic disk. Then, at some age (see Sect.~\ref{age}) its velocity increases either because of a dynamic interaction or because of a binary disruption. Thus, we have to specify two velocity distributions: before and after ejection.


The velocity components of newborn MS stars follow a normal distribution. Dispersions along the three axes of the velocity ellipsoid reflect random motions added to the general Galactic rotation. The velocity dispersion of OB stars in the Galactic coordinate system is characterized by the three principal velocity dispersions \citep{2022AstL...48..243B}:
\begin{equation}
\begin{aligned}
    \sigma_1 &= 14.98 \pm 0.08 ~\mathrm{km\, s^{-1}}, \\
    \sigma_2 &= 8.86  \pm 0.05 ~\mathrm{km\, s^{-1}}, \\ 
    \sigma_3 &= 7.58 \pm 0.04 ~\mathrm{km\, s^{-1}},
\end{aligned}
\end{equation}
that are not aligned with the Galactic coordinate axes $(U, V, W)$ but are instead rotated by specific angles:
\begin{equation}
\begin{aligned}
    L_1 &= 80.2 \pm 0.1^\circ, & B_1 &= 8.9 \pm 0.1^\circ, \\
    L_2 &= 170.2 \pm 0.1^\circ, & B_2 &= 0.1 \pm 0.1^\circ, \\
    L_3 &= 260.8 \pm 0.1^\circ, & B_3 &= 81.1 \pm 0.1^\circ.
\end{aligned}
\end{equation}
Each pair of $L$ and $B$ represents the direction of a principal axis in the Galactic coordinates. Since the velocity ellipsoid is tilted relative to the $(U, V, W)$ system, we need to apply a rotational transformation to express the velocity components in the Galactic velocity coordinates. To transform from the principal velocity frame to the Galactic velocity frame $(U, V, W)$, we use the rotation matrix $(R)$ using the unit vectors of the principal axes:
\begin{equation}
R =
\begin{bmatrix}
\cos L_1 \cos B_1 & \cos L_2 \cos B_2 & \cos L_3 \cos B_3 \\
\sin L_1 \cos B_1 & \sin L_2 \cos B_2 & \sin L_3 \cos B_3 \\
\sin B_1 & \sin B_2 & \sin B_3
\end{bmatrix}.
\end{equation}
The velocity dispersion tensor in the principal axes frame $\Sigma_{V}$ is given by:
\begin{equation}
\Sigma_{V} =
\begin{bmatrix}
\sigma_1^2 & 0 & 0 \\
0 & \sigma_2^2 & 0 \\
0 & 0 & \sigma_3^2
\end{bmatrix}
\end{equation}
which becomes in the Galactic velocity frame $(U, V, W)$
\begin{equation}
\Sigma_{UVW} = R \, \Sigma_V\, R^T.
\end{equation}
The initial random velocity components are sampled from the normal distribution with the given covariance matrix $\Sigma_{UVW}$ employing \textsc{galpy} \citep{2015galpy}. The circular velocity, dependent on $R$ and $z$ and also calculated employing \textsc{galpy}, is added to the initial random velocity. 

At the moment of ejection, the velocity is instantly increased.
As discussed in \cite{2011Silva}, the distribution of the (scalar) ejection velocity of runaway stars is bimodal with a maximum value of $400-500~\mathrm{km\,s^{-1}}$. 
This bimodality can be a natural consequence of BEM. 
However, since the difference between the two components is small, it can be assumed that the distribution of the scalar ejection velocity is described by a single Maxwellian that peaks at $156~\mathrm{km\,s^{-1}}$ \citep{2011Silva}. Ejection velocities are sampled from the Maxwellian distribution assuming isotropy, i.e. the direction is randomly drawn from a uniform distribution over the sphere, and the resulting velocity components are added to the one mentioned previously.

\subsubsection{Ages}\label{age}



The ejection time depends on the mechanism responsible for the ejection. Following previous studies \citep{2018ApJ...867L...8O, 2020ApJ...903...43D}, we assume that $2/3$ of the runaway stars are ejected by the DEM and $1/3$ by the BEM. As shown by \cite{2011Sci...334.1380F}, most Galactic runaway OB stars originate from star clusters that experience core collapse within the first $1~\mathrm{Myr}$ of their existence. Therefore, we assume that $2/3$ of the runaway stars are ejected $1~\mathrm{Myr}$ after birth. 

To determine the ejection time of a runaway star ejected via the BEM, we first select the mass of its binary companion and then compute its lifetime, which corresponds to the ejection time of the secondary star due to binary disruption produced by the SN explosion of the primary. The distribution of the mass ratio $q=M_2/M_1$ (with $M_2<M_1$) in binary stars is given by \citep{2012Sci...337..444S}:

\begin{equation}
    f(q)\sim q^{0.1}
    \label{eq:q}.
\end{equation}
Following common practice, $f(q)$ is approximated as a constant 
\citep{2024MNRAS.532.3926K}, meaning that the distribution is flat, so the initial masses of the primary and secondary components are independent of each other. First, we randomly find the primary component for every runaway star ejected by BEM, following the IMF given by Eq.~(\ref{eq:m>1}). Then, the lifetime of the primary component is calculated, and this corresponds to the time of ejection of the runaway star. 
Also, the lifetimes of all the runaway stars are calculated, which sets the time when they undergo an SN explosion.
Lifetimes for both the primary components and runaway stars are calculated by linear interpolation from the evolutionary track for metallicity $Z=0.01$ given in the \textsc{PARSEC} database\footnote{https://stev.oapd.inaf.it/PARSEC/} \citep{PARSEC}.

\subsection{Trajectories in the Galactic potential}
\label{galpy}

Following the procedure explained in Sect.~\ref{Initial parameters}, we obtain the masses, initial spatial distribution, velocity distribution from birth to ejection, velocity distribution after ejection, time of ejection depending on the ejection mechanism, and lifetime of runaway stars. After these parameters are set, the trajectories in the Galactic potential can be calculated. To do so, we use the \textsc{MWPotential2014} from the \textsc{galpy} library \citep{2015galpy}.

\textsc{MWPotential2014} is a realistic description of the Milky Way on small and large scales defined as:
\begin{itemize}
    \item the disk is modeled following the Miyamoto-Nagai approach  (\textsc{MiyamotoNagaiPotential}),
    \item the bulge is modeled with a power-law density profile that is exponentially cut-off with a power-law exponent of -1.8 and a cut-off radius of 1.9 kpc (\textsc{PowerSphericalPotentialCutoff}),
    \item the dark-matter halo is modeled following the Navarro–Frenk–White Potential (\textsc{NFWPotential}).
\end{itemize}
After integration, the spatial distribution of runaway OB stars in the Galaxy at the moment of SN explosion is obtained. 



\section{Results} \label{Results}

\subsection{OB stars} \label{OB}

\begin{figure}
    \resizebox{\hsize}{!}{\includegraphics{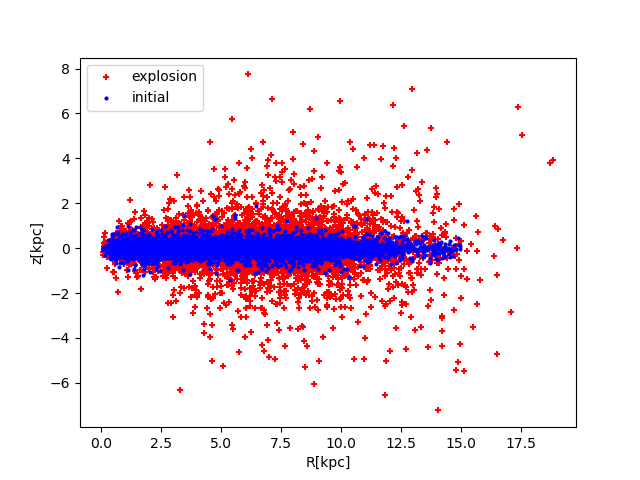}}
    \caption{The initial (blue dots) and final (red crosses) distribution of the runaway stars in the $R$-$z$ plane, where $R$ corresponds to the Galactocentric radial distance and $z$ to the height above the Galactic plane.}
    \label{fig:rz}
\end{figure}

In our sample, we use $N_\mathrm{stars}\approx60\,000$ stars with masses $8<M/M_\sun<55$ and $N_\mathrm{runaway}\approx7\,000$ of them are runaway stars.
In Fig.~\ref{fig:rz}, the initial and final distributions in the $R$-$z$ plane of the runaway stars are shown. It can be seen that, at the initial time, the stars are concentrated towards the Galactic disk, whereas at the time of the explosion, the dispersion around the Galactic plane is greater, and the stars can end up high above the Galactic plane, in the thick disk and in the Galactic halo. 

\begin{figure}
    \resizebox{\hsize}{!}{\includegraphics{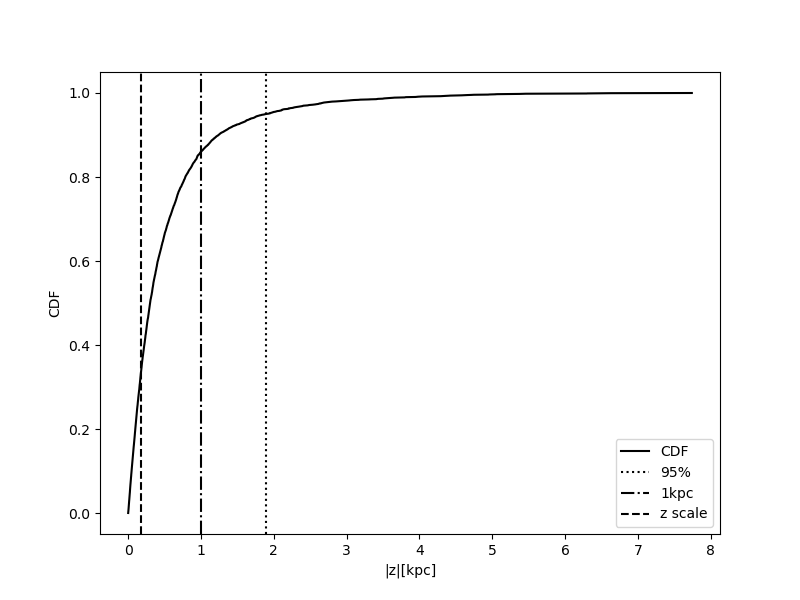}}
    \caption{The final cumulative distribution function of the height above the Galactic plane, $|z|$ (solid line). The vertical lines mark the scale height of the initial distribution (dashed line), $1\ \mathrm{kpc}$ (dash-dotted line) and the value of $z$ below which 95\% of runaway stars are concentrated (dotted line).}
    \label{fig:cdf}
\end{figure}

In Fig.~\ref{fig:cdf}, we present the cumulative distribution function (CDF) of the absolute value of the height above the Galactic plane $|z|$ at the moment of the SN explosion. Our model predicts that around 85\% of the runaway stars will explode below 1 kpc, and 95\%  at $|z|\lesssim1.9\ \mathrm{kpc}$, suggesting that a non-negligible fraction of runaway OB stars can explode in the Galactic halo.

The final distribution is calculated using a large number of runaway stars, on the order of $10^3$, to ensure statistically significant results. In order to estimate the real number of SNRs coming from runaway OB stars in the Galaxy, we follow two different approaches, either based on the star formation rate (SFR) or the CCSN rate.

First, we calculate the weighted fraction of runaway stars
\begin{equation}
    p = \frac{\int_{8M_\sun}^{55M_\sun}p(M)M^{-x}dM}{\int_{8M_\sun}^{55M_\sun} M^{-x}dM}.
    \label{eq:p}
\end{equation}
where $p(M)$ is the fraction of runaway stars obtained by linear interpolation of the values reported in Table~\ref{tab:ob}.

The SFR in our Galaxy is $\mathrm{SFR} = 2.0 \pm 0.7~M_\sun ~\mathrm{yr^{-1}}$ \citep{2022ApJ...941..162E}. 
The total number of stars with $0.08<M/M_\sun<150$ ($N_\mathrm{total}$) formed in the Galaxy during time $t$ (e.g. the average lifetime of an SNR)
is
\begin{equation}
    N_\mathrm{total} = \frac{\mathrm{SFR}\times t}{<M>},
\end{equation}
where $<M>$ is the average stellar mass, calculated using the Chabrier IMF (Eq.~\ref{eq:m<1} and Eq.~\ref{eq:m>1}):
\begin{equation}
    <M> = \frac{\int_{0.08M_\sun}^{150M_\sun}MdN}{\int_{0.08M_\sun}^{150M_\sun}dN}.
    \label{eq:meanm}
\end{equation}
The number of existing SNRs originating from runaway OB stars is then given by
\begin{equation}
    N_\mathrm{SFR} = N_\mathrm{total}\times  \frac{N_\mathrm{OB}}{N_\mathrm{total}}\times p,
    \label{eq:n-sfr}
\end{equation}
where the ratio $N_\mathrm{OB}/N_\mathrm{total}$ can be found integrating Eq.~\ref{eq:m<1} and Eq.~\ref{eq:m>1}.

On the other hand, the CCSN rate in our galaxy is estimated to be $\mathrm{CCSN} = 1.9 \pm 1.1 ~\mathrm{century^{-1}}$ \citep{2006Natur.439...45D}. The number of potentially observable SNRs originating from runaway OB stars, $N_\mathrm{CCSN}$, is obtained by multiplying the CCSN rate by the characteristic time and by the fraction of runaway stars
\begin{equation}
    N_\mathrm{{CCSN}} = t\times \mathrm{CCSN} \times p\,.
    \label{eq:n-ccsn}
\end{equation}

Using Eq.~\ref{eq:p}, we calculate $p=0.11$ and considering the typical lifetime of SNRs as $t=100~\mathrm{kyr}$, we find that the number of SNRs originating from runaway OB stars is around 310 when using the SFR and around 210 when using the CCSN rate. The two values are consistent within the uncertainties (based on errors in the CCSN,
$1.9 \pm 1.1 ~\mathrm{century^{-1}}$, and the SFR, $2.0 \pm 0.7~M_\sun ~\mathrm{yr^{-1}}$).

\subsection{Calvera}\label{Calvera}

As mentioned above, Calvera is a young NS with a characteristic age of $\tau_c = 285 ~\mathrm{kyr}$ \citep{2013ApJ...778..120H}. It is suspected that Calvera originated from an OB runaway star \citep{2008A&A...482..617P}. Here, we want to estimate the probability that a runaway OB star went supernova in a Galactic region characterized by Calvera's location.

Calvera is located at a distance of $\approx 3.3~\mathrm{kpc}$ and at a height above the Galactic plane of $\approx 2~\mathrm{kpc}$ \citep{2008ApJ...672.1137R}. Therefore, by fixing the values of $\bar r=3.3~\mathrm{kpc}$ and $\bar z=2~\mathrm{kpc}$, we find the probability density of an SN explosion in the cylinder with radius $\rho = \sqrt{\bar r^2-\bar z^2}$ and $|z|>\bar z$. Then, to obtain the probability, we multiply the probability density by the number of SN explosions coming from runaway OB stars $N_\mathrm{SFR}$ and $N_\mathrm{CCSN}$, obtained from Eq.~(\ref{eq:n-sfr}) and Eq.~(\ref{eq:n-ccsn}), using $t=\tau_c$. \\
We obtain $0.7-1.3$ using SFR and $0.5-0.9$ for CCSN. This suggests that there should be $\sim 1$ object within the cylinder originating from runaway OB star SN explosions in $285\ \mathrm{kyr}$, supporting the idea that Calvera could have originated from a runaway OB star. 

On the other hand, it was recently proposed that Calvera was created in the core-collapse following a SN explosion $<10~\mathrm{kyr}$ ago \cite[][]{2024ApJ...976..228R}. Again, using Eq.~(\ref{eq:n-sfr}), Eq.~(\ref{eq:n-ccsn}) and the upper limit of $t=10~\mathrm{kyr}$, we obtain $0.02-0.05$ and $0.02-0.03$ objects originating from the SN explosion of runaway OB stars in the cylinder in 10 kyr. Even in this case, the probability that Calvera originated from the SN explosion of a runaway star is not negligible. \\
The inverse transform sampling procedure introduces inherent statistical fluctuations between runs. Consequently, we report a range of values rather than a single deterministic outcome to account for this variability.

\subsection{SNRs}\label{snr}

In this subsection, we compare the probability of finding an SNR high above the Galactic plane originating from Type Ia SN \cite[see][Section 3.6]{2021Churazov} and from CCSN explosions of runaway OB stars. As explained in \cite{2021Churazov}, to obtain the number of SNRs originating from Type Ia SN explosions at $|z|\geq 1~\mathrm{kpc}$, it is necessary to analyze the rate of SN~Ia in the thick disk and the halo. As the Galaxy halo has a stellar mass of $\sim 1.4\times10^9M_\sun$ and a thick disk of $6\times10^9M_\sun$ \citep{2019MNRAS.490.3426D}, and adopting the observational estimate of the SN~Ia rate for elliptical galaxies with
stellar mass $<10^{11}M_\sun$ of $\sim 0.09$ SN~Ia per $100\ \mathrm{yr}$ per $10^{10}M_\sun$ \citep{2011MNRAS.412.1473L}, the SN~Ia rate of the Galaxy is $\sim 5.4\times10^{-4}$ per year in the thick disk and $\sim 1.3\times10^{-4}$  per year in the halo. To have the same proportionality and statistically significant results, we sampled $5\,400$ objects from the thick disk and $1\,300$ objects from the halo. 

The spatial density distribution of the objects sampled from the thick disk ($\rho_D$) follows the exponential law 
\begin{equation}
    \rho_D \propto \mathrm{exp} \left( -\frac{R}{h_R}\right)\mathrm{exp}\left(-\frac{z}{h_z}\right),
\end{equation}
where $h_R = 2.1~\mathrm{kpc}$ and $h_z = 0.9~\mathrm{kpc}$ are radial vertical and height scales \citep{2016ARA&A..54..529B}. The spatial density distribution of the objects sampled from the halo ($\rho_H$) underlay the spheroidal distribution along the galactocentric distance and a broken power-law: 
\begin{equation}
    \rho_H \propto (R^2+(z/0.6\,\mathrm{kpc})^2)^{-p/2},
\end{equation}
where $R$ and $z$ are in kpc,  $p = 2.3$ for $R\leq27 ~\mathrm{kpc}$, and $p = 4.6$ for $R\geq27 ~\mathrm{kpc}$ \citep{2011MNRAS.416.2903D}. 

After sampling SNRs originating from an SN~Ia explosion, we seek to compare the dominance of Type Ia and CC SNRs as a function of Galactic height, identifying the regions where each type prevails. As the total SN~Ia rate for $|z|\geq1~\mathrm{kpc}$ is $\sim 5.4\times10^{-4} + 1.3\times10^{-4} = 6.7\times10^{-4}$ SN~Ia per year, and the CCSN rate is $1.9\times10^{-2}$ per year, we see that there are around 28 times more SNRs originating from CCSN than from SN~Ia explosions. 

\begin{figure}
    \centering
    \resizebox{\hsize}{!}
    {\includegraphics{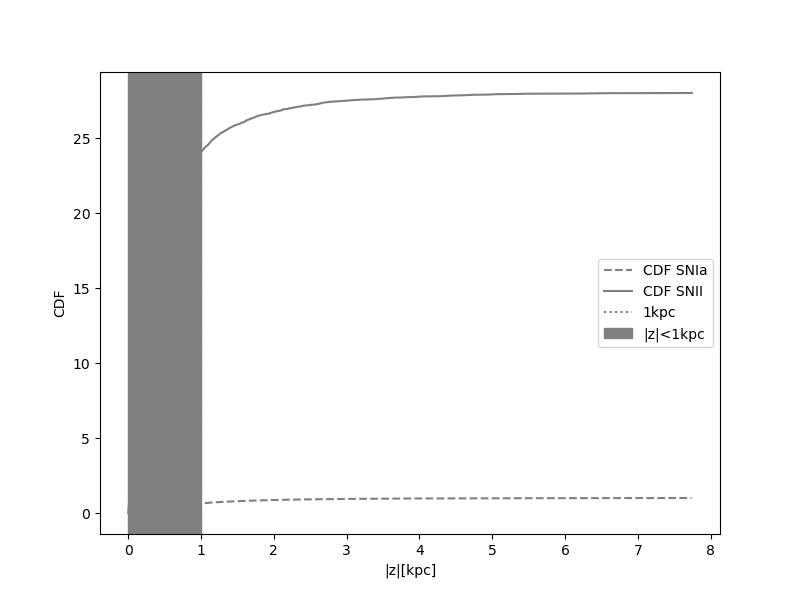}}
    \caption{The cumulative distribution function (CDF) of the absolute value of the height above the Galactic plane $|z|$ for SNRs originating from CCSN (solid line) and from SN~Ia (dashed line) with vertical line at $1~\mathrm{kpc}$ (dotted line). CDF of SNRs originating from CCSN is normalized to 28, as the rate of CCSN is 28x bigger than the one of SN~Ia. As SN~Ia from the thin disk are excluded from this calculation, only the difference for $|z|\geq1~\mathrm{kpc}$ should be observed.}
    \label{fig:SN~Ia}
\end{figure}

Therefore, we plotted the CDFs for SNRs originating from CCSN and SN~Ia depending on the height above the Galactic plane, where the CCSN one is normalized to 28 (Fig.~\ref{fig:SN~Ia}). It is important to note that the part of the figure for $|z|<1~\mathrm{kpc}$ is not relevant, as no SN~Ia in the thin disk was included in this calculation. In all $z$, the SNRs of the CCSN are dominant over those originating from SN~Ia.


\section{Discussion} 
\label{Discussion}

\subsection{OB stars}
\cite{2024Bisht} explored the possibility that the strong absorption lines from highly ionized gas can be explained if the line of sight passes through the reverse shocked material in a non-radiative SNR located outside the Galactic plane of the MW. 
In connection with this, the authors modeled the distribution of SNRs related to runaway OB stars.
It is therefore of interest to compare the results of the two approaches.

The results obtained in Sect.~\ref{OB} are slightly different from those presented in \cite{2024Bisht}. These authors have shown that around 93\% of the SNe explosions of runaway stars are confined within $|z|\leq1~\mathrm{kpc}$ and almost all within $|z|\leq2~\mathrm{kpc}$. In our model around 85\% are within $|z|\leq1~\mathrm{kpc}$ and around 95\% within $|z|\lesssim1.9~\mathrm{kpc}$. Unlike \cite{2024Bisht}, which used the uniformly distributed lifetime of runaway stars between $1-10~\mathrm{Myr}$ and the assumption that stars are ejected at birth, our approach of sampling stars from the IMF and calculating their lifetime based on masses from evolutionary tracks published in the \textsc{PARSEC} database allows for a more precise characterization of the final distribution of the SN explosion originating from runaway OB stars. We rely on simulation results for greater accuracy, since the approximate mass--lifetime relation \citep{2017imas.book.....C}
\begin{equation}
    t = 10^{10}\left(\frac{M_\sun}{M}\right) ^{2.5}\mathrm{yr}
\end{equation}
underestimates the lifetime of massive stars.
Moreover, we follow the motion before the ejection and determine the moment of ejection, which provides further improvements.
In our model, around $40\%$ of runaway stars are ejected more than $10~\mathrm{Myr}$ before they explode. As a result, stars can continue to move longer at their ejection velocities, which naturally explains their presence at larger $|z|$. 

Another key distinction is that we assume that the initial spatial distribution of OB stars follows that of young pulsars, whereas in \cite{2024Bisht} the initial radial coordinate of the birthplace is assumed to follow the stellar mass distribution.
Our choice is motivated by the fact that
the initial distribution of radio pulsars better traces the distribution of the progenitors in the last $\lesssim 10$~Myr. Another possibility is to assume that the distribution of OB stars follows that of SNRs, having the same form as Eq.~(\ref{eq:spatial}), but with different coefficients $\alpha$ and $\beta$ \citep{1998ApJ...504..761C}.
However, we believe that pulsars provide a better tracer because of their larger number of known objects and the more accurate distance determinations. Still, it is important to note that this model does not account for the spiral structure
of the Galaxy and could be refined with a more detailed model of the distribution of young massive stars. Despite this, the primary aim of this work is to investigate the distribution of CCSNe, which predominantly occur at intermediate galactocentric distances, and to explore a possible origin scenario for Calvera, which is relatively nearby. The model is not accurate in the central few kpc or in the outer regions beyond $\sim12~\mathrm{kpc}$. Given
that less than 5\% of the stars in our initial distribution lie beyond $12~\mathrm{kpc}$, this limitation has a minimal impact on the overall results. The results within the central few kpc should be interpreted with caution, as the model
does not adequately capture the complexity of this region.

Our model could be upgraded using the bimodal velocity distribution and Eq.~(\ref{eq:q}) instead of the approximation of constant $q$, but since these differences are tiny and due to our statistical approach, this would not result in significant differences. As mentioned before, the bimodality of the observed distribution is natural consequences of the BEM and we used the assumption that one third of stars are ejected by BEM, and the remaining ones by DEM. This assumption relies on the analyses of runaway stars in the SMC \citep{2018ApJ...867L...8O, 2020ApJ...903...43D}, and is in agreement with the results for the Galaxy \citep{2019A&A...624A..66R}. As shown in \cite{2019A&A...624A..66R}, BEM contributes to the population of unbound massive stars, and it predominantly produces walkaway stars rather than high-velocity runaways. \cite{2019A&A...624A..66R} suggest a runaway fraction among
stars more massive than $15~M_\sun$ between $0.2\%$ and $2.6\%$, and velocities mainly $\lesssim60~\mathrm{kms^{-1}}$. This is in contrast to the DEM, which is more efficient at producing high-speed runaway stars. In addition, they found that the runaway fraction and velocity depend on the metallicity, with five times more runaway stars with masses larger than $7.5~M_\sun$ for lower metallicity. Still, assuming that approximately 11\% of all OB stars are runaway stars (as it follows from Eq.~\ref{eq:p}), and that one third of these are produced by the BEM, implies that about $3.7\%$ of all OB stars become runaway stars via the BEM. This estimate is close to the values suggested by \cite{2019A&A...624A..66R}, who found that only $\sim 0.5\%$ of stars with $M\gtrsim15~M_\sun$ become high-velocity runaways via BEM, and around 5 times more when  $M\gtrsim7.5~M_\sun$. Although these two works support the dominance of the DEM for runaway OB stars, other studies, such as \cite{2022A&A...668L...5S}, suggest that BEM may play a significant role in certain cases, challenging a unified explanation. In addition, we do not account for the two-step ejection mechanism \citep{2010MNRAS.404.1564P}, nor for the SCES \citep{2024A&A...690A.207P}, as the first one might be a relatively rare process and the role and properties of the second one are not well-understood, yet. Therefore, further investigation of the ejection mechanisms, including their contribution to the total number of ejected stars and the velocities they impart, could provide valuable input for improving this model.

Furthermore, our upper mass limit of a runaway star is $55M_\sun$, due to interpolation, but we do not expect the final result to be sensitive to the upper mass limit, as more massive stars are rare and have a short lifetime, making it unlikely that they move to high $|z|$ considering the ejection velocity distribution. Also, the upper mass limit for a star to explode in the CCSN is still an open question, and recent studies, such as \cite{2012ARNPS..62..407J, 2018MNRAS.474.2116D, 2019MNRAS.490.4515S}, suggest
a value between $20-40M_\sun$. Still, the upper limit is highly dependent on the metallicity and, as discussed in \cite{2003ApJ...591..288H}, stars with higher metallicity could experience CCSN even when their initial masses are $>40M_\sun$. For these reasons, we think that our choice of $55M_\sun$ as the upper limit is reasonable.  
The final statistics are definitely more sensitive to the lower limit on the mass, which is set at $8M_\sun$ in our study. This depends on both metallicity and evolution. According to the simulations by \cite{2013ApJ...765L..43I}, the minimum mass is $\sim8.3M_\sun$, which is very close to the value we used.

The percentage of runaway OB stars as a function of spectral type and mass, presented in Table \ref{table:ob}, is obtained by linear interpolation. This approach may introduce some uncertainties, but, due to the limited number of bins, alternative interpolation methods are not suitable in this case. 


\subsection{Calvera and SNRs}

Our model predicts that, considering the characteristic age of Calvera as its true age, 
there is a high probability that the progenitor was a runaway star.
Therefore, Calvera could be explained in our model. However, there can be at most a few other NSs high above the Galactic plane with ages $\lesssim$~a few hundred thousand years at comparable distances or closer.

In a recent study \citep{2024ApJ...976..228R}, the kinematic age of Calvera was determined to be $<10~\mathrm{kyr}$. This is in correspondence with the age of the SNR G118.4+37.0, possibly related to Calvera \citep{2022A&A...667A..71A, 2023MNRAS.518.4132A}. 
In this case, the probability that Calvera originated from a runaway OB star is significantly lower but not negligible, about a few percent.

The results presented above in Sect.~\ref{Calvera} demonstrate that the probability that Calvera originated from a runaway star is greater for larger ages of the NS.
Although the true age of Calvera is expected to be lower than the characteristic age, it could not be as low as the kinematic age and the age of the associated SNR G118.4+37.0. As argued in \cite{2024ApJ...976..228R}, the kinematic age of Calvera could be affected by the uncertainties of the distance determined from its thermal X-ray emission \citep{2021ApJ...922..253M}. However, the age of the associated SNR, as discussed in \cite{2022A&A...667A..71A}, could be a poor estimate if the Sedov-Taylor self-similar solution is not appropriate, which is possible for SNRs expanding in a low-density medium such as that in the Galactic halo. Furthermore, as stated in \cite{2023MNRAS.518.4132A}, the age of the SNR was determined using the distance obtained for Calvera \citep{2021ApJ...922..253M}, which could add more uncertainties to the final result. Future improved age estimates of Calvera and of SNR G118.4+37.0 could lead to a more accurate and robust interpretation of our results.

Finally, our model predicts (see Sect.~\ref{snr}) that the SNRs observed at high Galactic latitudes could be explained better in terms of SN explosions of runaway OB stars than of SN~Ia. Since the velocity of runaway stars is limited to $\approx500~\mathrm{kms^{-1}}$ and the lifetime to $\approx50~\mathrm{Myr}$, they cannot explode above $|z|\approx25~\mathrm{kpc}$. Therefore, at very large $|z|$, only SNRs originating from SN~Ia could occur. 
This is important for the interpretation of the recently observed SNRs, e.g., presented by \cite{2015ApJ...812...37F} and \cite{2021Churazov}. As shown in \cite{2025Green}, during the last few years, SNRs have been discovered at unexpectedly high latitudes. Considering the results that we presented in Sect.~\ref{OB}, there could be more of them. However, at this moment, all the SNRs with a determined distance, and therefore known $z$, presented in the catalog by \cite{2025Green} are at $|z|<1~\mathrm{kpc}$.
New surveys for high-latitude SNRs are desirable. 

A similar comparative approach was taken by \cite{2021MNRAS.507..927M}, who examined the distributions of the CCSNe and SNe Ia distributions in the context of Galactic SNRs. However, their analysis was performed by static sampling of CCSNe in a thin disk and SN~Ia in a thin and thick disk, mostly corresponding to $|z|\lesssim 1~\mathrm{kpc}$. Despite the common goal of contrasting the two types of SNe, the scope and spatial domains of the two studies differ significantly. Our analysis emphasizes high-latitude regions, while theirs is restricted to $|b|\lesssim15^\circ$. These differences naturally lead to different interpretations and observational constraints.

 The number of SN explosions originating from runaway OB stars presented in Sect.~\ref{OB}, as well as the probability of finding an object like Calvera discussed in Sect.~\ref{Calvera}, depend mainly on the choice of IMF, the SFR and CCSN rate, Eq.~(\ref{eq:n-sfr}) and Eq.~(\ref{eq:n-ccsn}. The calculation was performed using both the Chabrier IMF and other formulations, such as the Miller-Scalo \citep{1979ApJS...41..513M} and Kroupa IMFs \citep{Kroupa02}. The results presented in Sect.~\ref{OB} show only minor variations when obtained using the Kroupa IMF. Such behavior is expected, as they have a similar power-law slope for $8<M/M_\sun<55$. When using the Miller-Scalo IMF, 95\% of stars will explode at $|z|\lesssim2.5~\mathrm{kpc}$, while the value of $|z|\lesssim1.9~\mathrm{kpc}$ is obtained using both the Kroupa and Chabrier IMF. This difference can be attributed to the steeper slope of the Miller-Scalo IMF, which leads to fewer massive stars being sampled. Consequently, a larger fraction of lower mass stars, characterized by longer lifetimes, are selected, allowing them more time to travel and potentially reach greater vertical distances ($|z|$) than those produced using the Kroupa or Chabrier IMFs.
 
Also, the number of SN explosions in $t=100~\mathrm{kyr}$ originating from runaway OB stars and the probability of finding an object like Calvera depend on the choice of the IMF (Sects.~\ref{OB} and \ref{Calvera}). When using Kroupa IMF, the number of SNRs in $t=100~\mathrm{kyr}$ originating from runaway OB stars is around 190 when using SFR and around 210 when using the CCSN rate. In both cases, the number of Calvera-like objects is between 0.4 and 0.8, again suggesting that there could be a $\sim1$ object within the cylinder originating from runaway OB star SN explosions in $285~\mathrm{kyr}$. Using $t = 10~\mathrm{kyr}$, we obtain 0.01 - 0.02 objects originating from the SN explosion of runaway OB stars in the cylinder in $10~\mathrm{kyr}$. On the other hand, when adopting the Miller-Scalo IMF, the number of SNRs originating from runaway OB stars is around 110 when using the SFR and around 160 when using the CCSN rate. Therefore, the probability based on SFR is 0.4 - 0.9, while the CCSN-based estimate is 0.6 - 1.4, again suggesting that there could be a $\sim1$ object within the cylinder originating from runaway OB star SN explosions in $285~\mathrm{kyr}$. Using $t = 10~\mathrm{kyr}$, we obtain 0.01 - 0.03 (SFR) and 0.02 - 0.05 (CCSN) objects originating from the SN explosion of runaway OB stars in the cylinder in $10~\mathrm{kyr}$. Here we can see that, even though the final results depend on the choice of the IMF, the differences are tiny. 
 
 The value we used for SFR is based on data from the Herschel Infrared Galactic Plane Survey \citep{2022ApJ...941..162E}. The presented value of SFR is in agreement with all previous studies, except one \cite[see][Table 1]{2022ApJ...941..162E}.
The CCSN rate, adopted from \cite{2006Natur.439...45D}, is estimated by analyzing gamma rays from radioactive $^{26}\mathrm{Al}$, which is in agreement with other studies, e.g. \cite{1993A&A...273..383C, 2005AJ....130.1652R, 2021NewA...8301498R}. However, recently lower estimates of the Galactic SFR and the corresponding CCSN rate have been proposed. 

In the recent study by \cite{2025MNRAS.538.1367Q}, the Galactic SFR and CCSN rates are derived from the census of OB stars within $1~\mathrm{kpc}$ of the Sun. The value of Galactic SFR, $\mathrm{SFR}=0.67^{+0.09}_{-0.01}M_\sun\mathrm{yr^{-1}}$, is in agreement with that presented in \cite{2010ApJ...710L..11R}, but is significantly lower compared to many earlier studies \cite[see][Table 1]{2022ApJ...941..162E}. The authors relate this difference to a more complete OB star catalog that they use. 
In addition, the estimated value of the Galactic CCSN rate, $\mathrm{CCSN} = 0.4\pm0.1~\mathrm{century^{-1}}$  (with an upper mass limit on CCSN progenitors) and $\mathrm{CCSN} = 0.5\pm0.1~\mathrm{century^{-1}}$ (without an upper mass limit on CCSN progenitors), is lower than the previous estimates, for example \cite{1993A&A...273..383C, 2005AJ....130.1652R, 2006Natur.439...45D, 2021NewA...8301498R}. As claimed in this work, the values are significantly lower primarily because of the smaller size of the OB catalog combined with improved stellar evolutionary models. 

Using the values of SFR and CCSN rate without an upper mass limit on CCSN progenitors  \cite[][]{2025MNRAS.538.1367Q} in Eq.~(\ref{eq:n-sfr}) and Eq.~(\ref{eq:n-ccsn}) and the Chabrier IMF, we find that the number of SNRs originating from runaway OB stars is around 100 (SFR) and around 45 (CCSN rate). Naturally, these results are significantly lower than those presented in Sect.~\ref{OB}. In the case of Calvera, using its characteristic age, we expect $0.2-0.4$ (SFR) and $0.2-0.3$ (CCSN rate) of such objects in the volume considered, and using its kinematic age, we expect only $0.007-0.014$ (SFR) and $0.007-0.010$ (CCSN rate). In this case, the probability that Calvera originated from a runaway OB star is very low. This brings us back to the uncertainties mentioned previously in age estimates.

\section{Conclusions} 
\label{Conclusion}

In this study, we have analyzed the spatial distribution and rate of runaway OB stars relative to the Galactic plane, and the key conclusions are summarized below. 

\begin{itemize}
    \item Around $85\%$ of the runaway OB stars will explode within $|z|\leq1~\mathrm{kpc}$ and around $95\%$ within $|z|\lesssim1.9~\mathrm{kpc}$. This suggests that a significant fraction of runaway OB stars may undergo SN explosions in the Galactic halo.
    \item We expect to have around 310 SNRs (using the SFR) and around 210 SNRs (using the CCSN rate) with ages $<10^5$~yrs originating from runaway OB stars. These numbers depend mainly on the SFR and CCSN rate. Considering the latest proposed lower estimates, we expect to have around 100 SNRs (using the SFR) and around 45 SNRs (using the CCSN rate) younger than $10^5$~yrs.
    \item There is a possibility to explain young NS Calvera using our model, and the probability is greater for a larger age of Calvera. These results depend on both the SFR and the CCSN rate, as well as on the estimated age of Calvera and the associated SNR G118.4+37.0.
    \item We have compared the probabilities of finding an SNR originating from CCSN high above the Galactic plane with those originating from SN~Ia. The SNRs from CCSNe predominate over those resulting from SN~Ia. However, since runaway stars have limited velocity, they cannot explode at very large distances from the Galactic plane. Therefore, at extreme heights, only SNRs originating from SN~Ia are expected.
\end{itemize}

These results support the hypothesis that Calvera originates from a runaway OB star and are also important for further analysis of the SNRs recently observed high above the Galactic plane. Future investigations in estimating the age of Calvera and the associated SNR G118.4+37.0, as well as in refining the Galactic SFR and the Galactic CCSN rate, are essential for accurately interpreting our results. Finally, new surveys are needed for high-latitude SNRs to validate our predicted numbers and distribution.

\begin{acknowledgements}
The authors thank the referee for useful comments and suggestions.
This work is based on a thesis submitted by VD in partial fulfillment of an MSc degree in Astrophysics and Cosmology at the University of Padova. 
\end{acknowledgements}

%
%

\bibliographystyle{aa} 
\bibliography{bib} 

\end{document}